\documentclass[preprint,12pt]{elsarticle}

\usepackage{graphicx}
\usepackage{float}
\DeclareGraphicsExtensions{.pdf,.png,.jpg}

\usepackage{amssymb}

\journal{Physical Letters A}

\begin{document}

\begin{frontmatter}



\title{Radiation pressure excitation of test mass ultrasonic modes via three mode opto-acoustic interactions in a suspended Fabry-$\rm P\acute{e}rot$ cavity}


\author{Carl Blair, Sunil Susmithian, Chunnong Zhao, Fang Qi, Li Ju, and David Blair}

\address{School of Physics, University of Western Australia, 35 Stirling Highway, Crawley, Western Australia 6009, Australia}

\begin{abstract}
Three-mode parametric-instabilities risk stable operation of gravitational-wave detectors. Instabilities occur through time
 varying radiation-pressure distributions, derived from beating between two optical modes, exciting mirror acoustic modes in
 Fabry-$\rm P\acute{e}rot$ cavities. Here we report the first demonstration of radiation-pressure driving of ultrasonic-acoustic
 modes via pairs of optical modes in gravitational-wave type optical cavities. In this experiment $\backsim0.4W$ of
 TEM$_{01}$ mode and $\backsim1kW$ of TEM$_{00}$ mode circulated inside the cavity, an $\backsim181.6kHz$
 excitation was observed with amplitude $\backsim5\times10^{-13}m$. The results verify the driving force term in the
 parametric instability feedback model\cite{brag01}. The interaction parametric gain was $(3.8\pm0.5)\times10^{-3}$ and mass-ratio scaled opto-acoustic overlap $2.7\pm0.4$.
\end{abstract}

\begin{keyword}
Parametric Instability \sep Three mode interaction \sep Gravitational waves

\end{keyword}

\end{frontmatter}


\section{Introduction}
\label{introduction}
Advanced laser interferometer gravitational wave detectors now under construction \cite{harry} \cite{virgo} aim to achieve a strain sensitivity $h\backsim 10^{-23}/ \sqrt{Hz}$ at $\backsim100Hz$. To achieve this sensitivity without recourse to quantum squeezing, it is necessary to balance the shot noise and the radiation pressure noise. To achieve this balance at $100Hz$ requires the optical power in the arm cavities to approach $1MW$. These conditions are also such that three mode parametric interactions can become strong, giving rise to parametric instability \cite{brag02}. Such instability arises when the thermal noise fluctuations in the surface profile of a cavity mirror (for a particular high quality factor mirror acoustic mode), scatters carrier light into a transverse optical mode. The scattered mode in turn beats with the fundamental cavity mode. The beating creates a time varying radiation pressure force distribution on the mirror. If the force distribution overlaps the acoustic mode shape, and if the transverse mode is resonant in the cavity and has the appropriate phase, the radiation pressure force acts in phase with the thermal fluctuations, thereby providing a positive feedback. If this feedback force is large enough, it can overcome the acoustic losses of the mode, leading to exponential growth of the acoustic mode. The Stokes interaction, for which the cavity transverse mode is lower than the carrier frequency, satisfies the phase requirement. 

\graphicspath{{./}}
\begin{figure} [!b]
\centering
\includegraphics[width=0.5\linewidth]{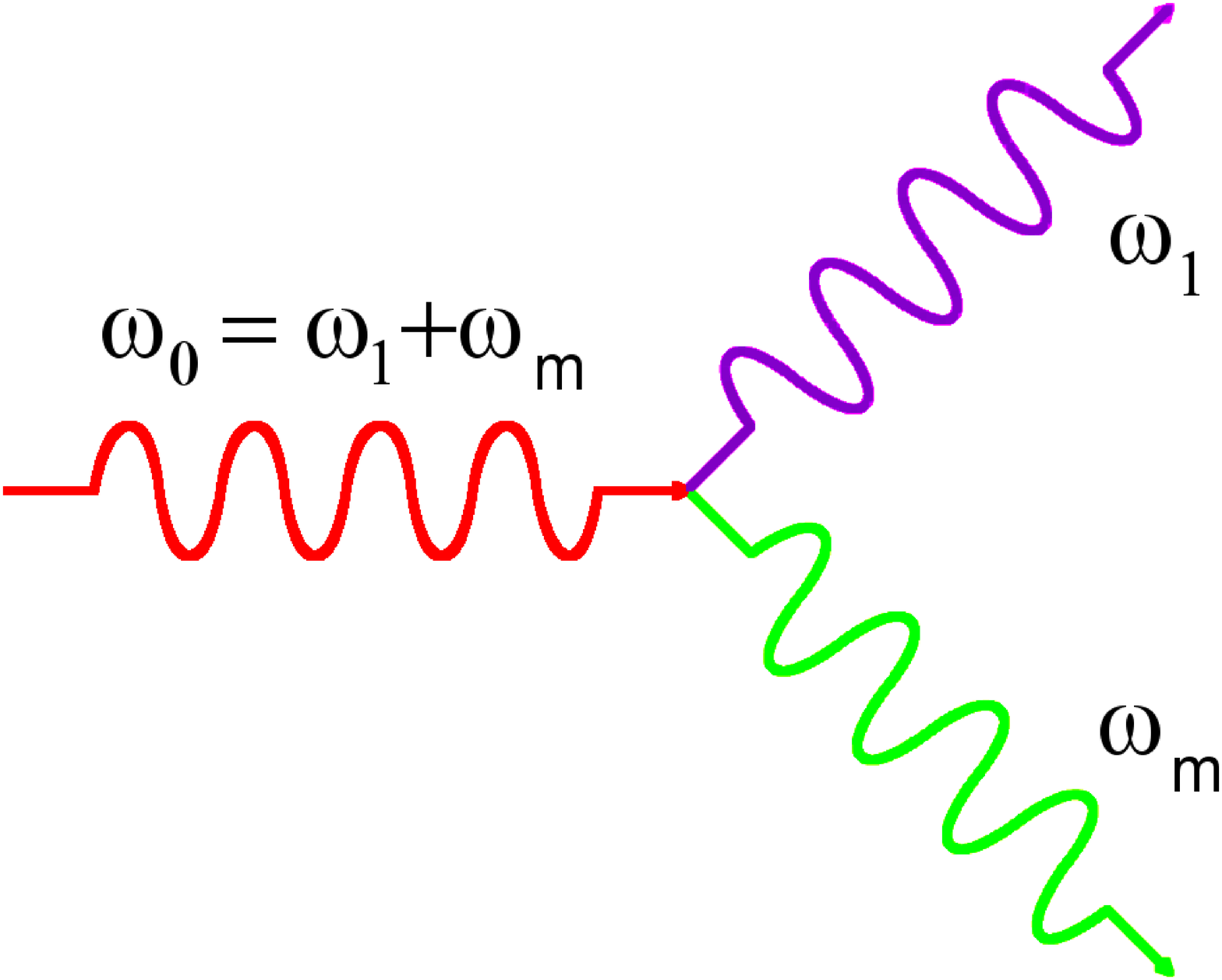}
\caption{Three-mode interaction. TEM$_{00}$ undergoes pair creation, to produce a phonon in the mirror and a resonant TEM$_{01}$ photon. }
\label{figure 1}
\end{figure}

It is instructive to consider the interaction in the quantum picture as illustrated by the Feynman diagram in figure \ref{figure 1}. The high occupation number of the carrier ensures that the 3-mode interaction (represented by a 3-quanta vertex) is dominated by photon-phonon pair creation. A carrier photon decays coherently into a transverse mode photon and a phonon. The phonons contribute to the occupation number of the acoustic mode, and if the rate of this process exceeds the thermal relaxation rate of the mode, the acoustic mode occupation number will increase with time. The process is only efficient for opto-acoustic systems that satisfy a triple resonance condition, and even when resonant, high optical power is required to achieve instability in gravitational wave detectors. Triple resonance means that the carrier, the transverse mode and the acoustic mode must satisfy $\omega _0 =\omega _1 +\omega _m$ where $\omega _0$ is the carrier frequency, $\omega _1$ is the transverse mode frequency and $\omega _m$ is the acoustic mode frequency.

Braginsky et al. \cite{brag02} derived the parametric gain $R$ (see equation \ref{ParaGain}) of the parametric 
instability process from a classical viewpoint. If the parametric gain exceeds unity for a 
particular acoustic mode, the mode will grow exponentially. The experiment described here is 
part of a research program aiming to determine the best way of controlling instability. 
\begin{equation}\label{ParaGain} R = \frac{4PQ_m}{McL\omega_m^2} \bigg(\frac{Q_s\Lambda_s}{1+(\Delta\omega_s/\gamma_s)^2}-\frac{Q_a\Lambda_a}{1+(\Delta\omega_a/\gamma_a)^2}\bigg) \end{equation}

Here $P$ is the stored power in the cavity, $m$ represents the mechanical oscillator, subscripts $s$ and 
$a$ denote the Stokes and anti-Stokes modes, $Q$ represents the quality factors of the mechanical 
and optical modes, $M$ is the mass of the test mass, $L$ is the length of the cavity, $\gamma$ represent the 
optical mode relaxation rates and $\Lambda$ the overlap factors (which include the mass ratio between the 
test mass total mass $M$ and the mechanical oscillator effective mass). 

In the configuration of current experiment (near flat-flat cavity), only the anti-Stokes process satisfies the resonance 
condition. A miniature system with the same configuration might find application in macroscopic 
quantum experiments \cite{zhao09}, in which the triple resonance reduces the power requirement 
significantly for reaching the quantum noise limit \cite{dobrindt}. 
Three-mode parametric instability can be broken down into components in the feedback loop 
model \cite{evans}. Two of these have been verified: a) Thermal tuning: It was predicted in 2005 that PI 
can be tuned through thermally induced radius of curvature variations in the cavity mirrors \cite{zhao05}. 
This tuning of the resonant interaction was demonstrated in 2008 \cite{miao}. b) Acoustic generation of 
transverse optical modes: The acoustic generation of a specific resonant transverse mode has 
amplitude determined by the overlap parameter – a number that determines the overlap between 
the acoustic and optical modes. This was reported by Zhao et al \cite{zhao08}. In 2011 thermally excited 
transverse modes were detected, and three mode interactions were shown to enable high 
sensitivity spectroscopy of thermally excited acoustic modes \cite{zhao11}. 
The third component of three mode interactions is the one that closes the feedback loop: 
radiation pressure excitation via the beating of the carrier with a low amplitude transverse mode. 
Here we report the first observation of such radiation pressure excitation in a long optical cavity 
designed as a sub-scale version of a gravitational wave detector cavity. The specific interaction 
being studied is the three mode interaction between a TEM$_{00}$ mode, a TEM$_{01}$
 mode tuned about $181.6kHz$ above the TEM$_{00}$ and a mirror acoustic mode at $181.6 kHz$. The system is shown 
schematically in figure \ref{experimental setup}. The cavity power is only about $1kW$, and spontaneous parametric 
instability cannot be achieved. However by artificially increasing the transverse mode power we 
can mimic the conditions of instability, and measure the excitation of the acoustic mode by 
observing its free ring down. This mechanism would result in instability if the parametric gain $R$ 
was greater than unity \cite{brag01}. 
In the experiment described here we used the anti-Stokes process which normally induces 
cooling rather than instability. The first term inside the bracket of equation \ref{ParaGain} is negligible because of $\Delta\omega_s$ is very large. However, in our experiment the parametric gain $R<<1$. To mimic the high 
gain regime we artificially generate TEM$_{01}$ power. Our results are dominated by the inserted 
TEM$_{01}$ power. In this situation the parametric gain has little effect on the results, although in a 
higher power experiment the negative parametric gain would lead to suppression of the driving 
term. In this experiment the use of the anti-Stokes process to demonstrate the radiation pressure 
excitation does not lead to loss of generality.

\section{Experimental Setup}
\label{experimental setup}
\begin{figure} [!h]
\centering
\includegraphics[width=0.85\linewidth, ]{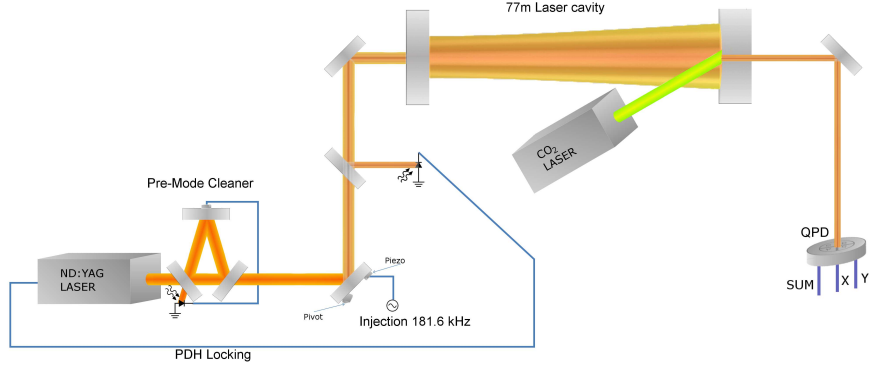}
\caption{Experimental Layout, showing laser, pre-mode cleaner, mechanical excitation of ~TEM$_{01}$
 using a piezo mirror, the PHD locking loop, the laser cavity with end mirror radius of curvature tuned using CO$_2$ 
laser heating and readout of the transmitted beam using a quadrant photo detector (QPD). }
\label{figure 2}
\end{figure}

The experimental layout is shown in Figure \ref{figure 2}. The Nd:YAG laser (1064$nm$) is frequency locked 
to the 80 $meter$ suspended cavity. The laser beam first passes through a pre-mode cleaner (PMC) 
to create a pure TEM$_{00}$ mode. The beam from the PMC is reflected by a piezo actuated mirror. 
The actuator is driven at a frequency matched to the mechanical resonant mode of the end 
test mass (ETM) mirror to generate TEM$_{01}$ mode modulation sidebands. The TEM$_{00}$ mode 
carrier with TEM$_{01}$ mode sidebands are injected into the 80 $meter$ suspended cavity. The CO$_2$ 
laser power is adjusted to tune the radius of curvature of the ETM until the cavity mode 
frequency gap between TEM$_{01}$ mode and TEM$_{00}$ mode matches the ETM acoustic mode 
frequency. This causes the TEM$_{00}$ carrier and a single TEM$_{01}$ sideband to resonate 
simultaneously inside the cavity. The beat frequency between the TEM$_{00}$ and TEM$_{01}$ is precisely tuned 
to the resonant frequency of the acoustic mode of interest of the ETM, at $181.6 kHz$. The 
transmitted beam is monitored using the quadrant photo-detector (QPD). When the TEM$_{01}$
is being injected, the QPD mainly measures the amplitude of the transmitted TEM$_{01}$ mode. 
After the driving is stopped, the QPD measures the TEM$_{01}$ mode amplitude generated by 
acoustic mode scattering, and therefore the amplitude of the acoustic mode. The measured ring-down curve 
of the signal after the excitation is switched off proves the radiation pressure 
excitation of the acoustic mode. 

\section{Experimental Results}
\label{experimental results}

Assuming that the injected TEM$_{00}$ mode and TEM$_{01}$ mode amplitude inside the cavity are $E_{00}$ 
and $E_{01}$, the excitation drives the acoustic mode amplitude to $X_m$ that scatters a part of the TEM$_{00}$
 mode into TEM$_{01}$ mode with amplitude of $E_{01}^2$ . If the power transmission of the ETM is $T$, the 
QPD detection efficiency is $g$, then the output of the QPD at the acoustic mode frequency and the 
DC component are;
\begin{equation}\label{driving} V_{drv} = 2gT | E_{00}(E_{01} + E_{01}^S) | \end{equation}
\begin{equation}\label{dcsig} V_{DC} = gT | E_{00} |^2 \end{equation}

\begin{figure} [!h]
\centering
\includegraphics[width=0.75\linewidth, ]{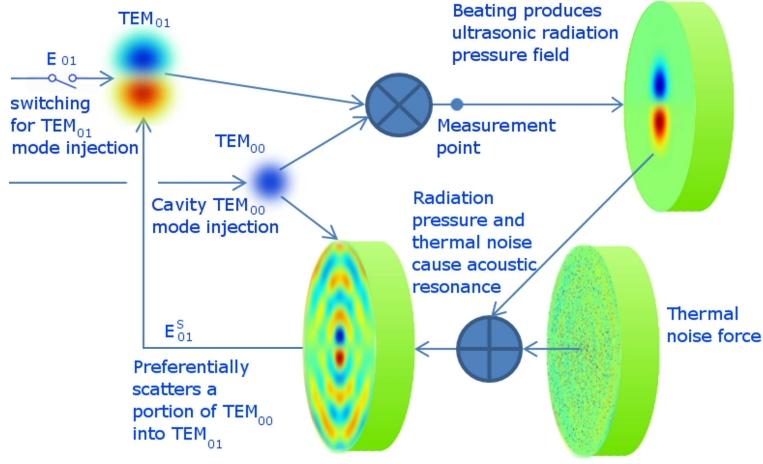}
\caption{Schematic block diagram of the three-mode opto-acoustic feedback loop. Both TEM$_{00}$ and TEM$_{01}$ 
modes are injected into the cavity. The beating between these two signals is detected at the measurement 
point, and provides the radiation pressure field on the mirror. The radiation pressure field and thermal 
noise both excite acoustic mode resonances of the mirror. When the injected TEM$_{01}$ mode is switched off 
the exponentially decaying beat signal demonstrates the radiation pressure excitation. }
\label{figure 3}
\end{figure}

Immediately after turning off the driving $(E_{01}=0)$, the output of the QPD at the acoustic mode frequency is:
\begin{equation}\label{readsig} V_{rd} = 2gT | E_{00}E_{01}^S | \end{equation}

Here $E_{01}^S$ is proportional to the acoustic mode amplitude and should ring down as the acoustic 
mode rings down. In the case of current experiment, the parametric gain $R$ is much smaller than 
unity. Hence the TEM$_{01}$ created by scattering should be much smaller than the driving field, i.e: $E_{01}^S<<E_{01}$. The parametric gain in the feedback loop model is the open loop gain. If we break the loop at the TEM$_{01}$ injection point and measure $E_{01}^S$ relative to $E_{01}$, then the parametric gain is 
given by: 
\begin{equation}\label{readsig} |R| = 2gT \bigg|\frac{E_{00}E_{01}^S}{E_{01}}\bigg| \approx \bigg|\frac{V_{rd}}{V_{drv}}\bigg| \end{equation}

Here we define time zero as the time that the driving signal is stopped. By measuring the ring 
down time we obtain the Q-factor of the acoustic mode. Note that when there is no injection of the driving field, the thermal excitation is the dominate source of the acoustic mode vibration. If the parametric gain was large, it would act to damp the ring-down signal. However for this experiment the low value of $R$ means that this effect is very small. 
 
\begin{figure} [H]
\centering
\includegraphics[width=8cm, ]{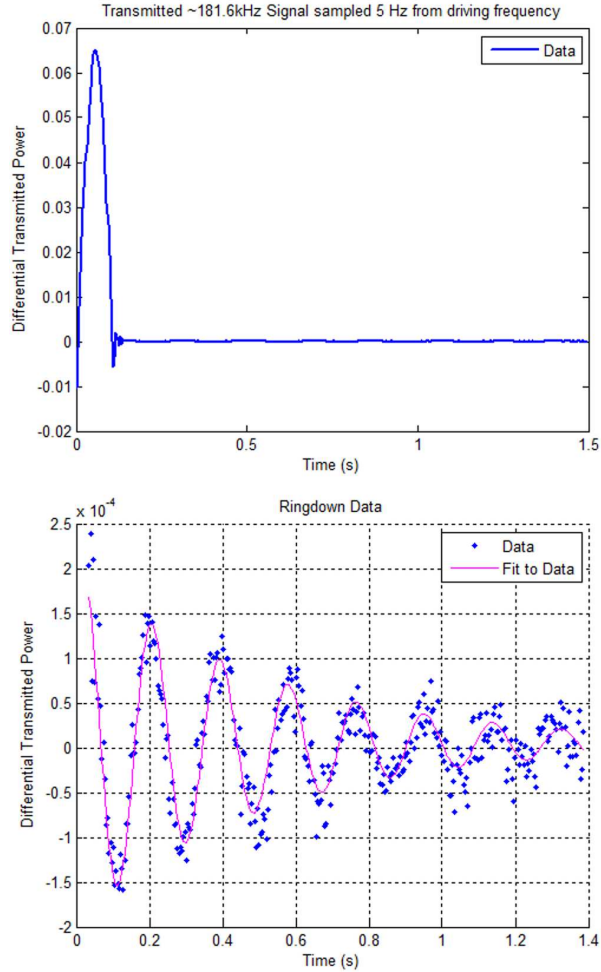}
\caption{Raw data, showing the moment the driving was switched off (top) and zoomed in to view the ring down (bottom) }
\label{figure 4}
\end{figure}

Figure \ref{figure 4} shows the output of the QPD y axis. Initially, the injection of the driving TEM$_{01}$
 mode field transmitted through the cavity, beats with the TEM$_{00}$ mode field and creates a high 
amplitude output at $181.6 kHz$. The spectrum analyzer is used to down-convert the signal 
frequency to $~5Hz$ for easy detection, with amplitude of $~64\pm 0.5mV$. In the top curve of figure 
\ref{figure 4}, the driving applied to the PZT was turned off at $0.109 seconds$ (top). The QPD output then 
can only be seen in the zoomed in graph (bottom curve figure \ref{figure 4}) because of the small amplitude. 
In this figure the reference time (the moment the driving was turned off) has been set to zero. 
The measurement presented here could be affected by the cavity ring down time and by 
electromagnetic transients. For this reason the first 0.035 seconds are not used in the analysis. 
We can clearly see the ring-down curve after turning off the driving. The curve fitting of the 
ring-down curve shows the ring-down time of $\approx0.5\pm 0.1 seconds$, corresponding to a mechanical 
Q-factor of $Q_m \approx (3 \pm 1)\times 10^5$. The ring-down amplitude at the time the driving was turned off was 
$V_{rd}(0)=0.24\pm 0.03 mV$. Using equation \ref{readsig} we use this result to determine the parametric gain $R 
\approx (3.8\pm 0.5) \times 10^{-3}$. Using equation \ref{ParaGain} we can now determine the mass scaled overlap factor to be , $2.7\pm 0.4$, using the effective mass ratio of $\approx 13$ from the ANSYS simulations, this can be translated 
to a naught to unity bounded overlap factor of approximately 0.21. The following parameters were used in the experiment:
\begin{equation}\label{parameters} M=5.6kg;P=10^3W;L=77m;\omega_m=2\pi\times181.6\times10^3Hz; Q_m=(3\pm 1)\times 10^5; Q_s = 2\times 10^{11}\end{equation}

The overlap parameter derived experimentally is about 51\% of the expected value for perfect 
alignment of the optical and acoustic modes used in this experiment. In multiple experiments we 
observed high sensitivity to alignment. This was expected because the overlap parameter is 
strongly modulated by lateral displacement of the optical modes relative to the test mass acoustic 
modes as investigated by Heinert et al \cite{heinert}. This also points to a simple method of reducing the 
parametric gain of a single mode near the threshold of instability, by finely adjusting the spot 
positions on the mirror test masses. 

\section{Conclusion}
\label{conclusion}
By injecting a high order mode simultaneously with the fundamental mode into an optical cavity 
we have demonstrated the radiation pressure driving term in the theory of three mode parametric 
instability. We have demonstrated a simple method for measuring the parametric gain and the 
overlap factor of three mode interactions at relatively low optical power. The technique can be 
used in long baseline interferometer gravitational wave detectors to identify and characterize 
potentially unstable acoustic modes in situ when the detector is operating at low optical power. 
This can allow the design of appropriate control schemes that will mitigate instabilities that will 
occur when sensitivity demands require operation at high optical power. We have also shown 
that parametric gain can be tuned by adjusting the laser beam spot positions so as to change the 
opto-acoustic overlap parameter.

\section{Acknowledgements}
\label{acknowledgements}
The authors gratefully acknowledge the support of the LIGO Scientific Collaboration, and 
especially the optics working group and the LSC Gingin Advisory Committee. The authors also 
gratefully acknowledge the Australian Consortium for Gravitational Astronomy, the Australian 
Research Council, and the Western Australian Centre of Excellence in Science and Innovation 
program.




\bibliographystyle{elsarticle-num}
\bibliography{<your-bib-database>}

\begin{thebibliography}{10}


\bibitem{brag01} V.B. Braginsky et al,{\it Phys Lett. A} {\bf 287} (2001) 311.
\bibitem{harry} G.M. Harry,{\it Classical and Quantum Gravity}, {\bf27-8} (2012) 084006.
\bibitem{virgo} The VIRGO project Web site, { http://www.virgo.infn.it/} Nov 2012.
\bibitem{brag02} V.B. Braginsky, S. Strigin, S. Vyatchanin, {\it Phys. Lett. A}{\bf 293} (2002) 228.
\bibitem{zhao09} C. Zhao, at al. {\it Phys. Rev. Lett.} {\bf 102} (2009) 243902 
\bibitem{dobrindt} J. M. Dobrindt and T. J. Kippenberg, {\it Phys. Rev. Lett.} {\bf 104}, (2010) 033901 
\bibitem{evans} M. Evans, et al., {\it Phys. Lett. A} {\bf 344} (2010) 4-665
\bibitem{zhao05} C. Zhao, L. Ju, J. Degallaix, et al., {\it Phys. Rev. Lett.} {\bf 94} (2005) 12110
\bibitem{miao} H. Miao, C. Zhao, L. Ju, et al., {\it Phys. Rev. A} {\bf 78} (2008) 063809. 
\bibitem{zhao08} C. Zhao, et al., {\it Phys. Rev. A} {\bf 78-2} (2008) 023807. 
\bibitem{zhao11} C. Zhao, Q. Fang, S. Susmithan, et al., {\it Phys. Rev. A}, {\bf 84} (2011) 06
\bibitem{heinert} D. Heinert,S.E. Strigin, {\it Phys. Lett. A}, {\bf 375-43} (2011) 3804-3810 

\end{thebibliography}



\end{document}